\documentclass[twocolumn]{aastex62}
\usepackage{natbib}
\graphicspath{{./}{figures/}}
\usepackage{color,subfigure,lineno} 
\newcommand{\hbeta}{H{$\beta$}}

\def\MgII{Mg\,{\sc ii}}

\def\OIII{[O\,{\sc iii}]\,5007}

\shorttitle{Quasar Episodes}
\shortauthors{Shen}
\begin{document}

%\title{A Statistical Constraint on the Episodic Quasar Lifetime}
\title{Extreme Variability and Episodic Lifetime of Quasars}

%\correspondingauthor{}
%\email{}

\author[0000-0003-1659-7035]{Yue Shen}
\affiliation{Department of Astronomy, University of Illinois at Urbana-Champaign, Urbana, IL 61801, USA}
\affiliation{National Center for Supercomputing Applications, University of Illinois at Urbana-Champaign, Urbana, IL 61801, USA}

\begin{abstract}
We constrain the average episodic quasar lifetime (as in steady-state accretion) using two statistics of quasars that are recently turned off (i.e., dimmed by a large factor): 1) the fraction of turned-off quasars in a statistical sample photometrically observed over an extended period (e.g., $\Delta t=20$ yrs); 2) the fraction of massive galaxies that show ``orphan'' broad \MgII\ emission, argued to be short-lived echoes of recently turned-off quasars. The two statistics constrain the average episodic quasar lifetime to be hundreds to thousands of years. Much longer (or shorter) episodic lifetimes are strongly disfavored by these observations. This average episodic lifetime is broadly consistent with the infall timescale (viscous time) in the standard accretion disk model for quasars, suggesting that quasar episodes are governed by accretion disk physics rather than by the gas supply on much larger scales. Compared with the cumulative quasar lifetime of $\sim 10^6-10^8\,$yrs constrained from quasar clustering and massive black hole demographics, our results suggest that there are $\sim 10^3-10^5$ episodes of quasar accretion during the assembly history of the supermassive black hole. Such short episodes should be clustered over intervals of $\sim 10^4\,$yrs to account for the sizes of ionized narrow-line regions in quasars. Our statistical argument also dictates that there will always be a small fraction of extreme variability quasars caught in ``state transitions'' over multi-year observing windows, despite the much longer episodic lifetime. These transitions could occur in a rather abrupt fashion during non-steady accretion.
\end{abstract}


\section{Introduction}\label{sec:introduction}

Quasars are efficiently accreting supermassive black holes (SMBHs) in galactic nuclei during which they assemble most of their present-day masses \citep{Soltan_1982,Yu_Tremaine_2002, Shankar_etal_2009}. The lifetime of quasars is an important quantity to understand the growth of SMBHs, and their roles in galaxy evolution. Measurements of quasar clustering \citep[e.g.,][]{Shen_etal_2007,Eftekharzadeh_etal_2015} as well as population synthesis of SMBH accretion \citep[e.g.,][]{Yu_Tremaine_2002,Shankar_etal_2009} have constrained the cumulative lifetime of quasars to be broadly within $10^6-10^8$\,yrs; see additional constraints on quasar lifetimes in \citet{Martini_2004}. However, it is unclear if quasar accretion is continuous over this time span, or episodically through many cycles of efficient accretion. Currently, lifetime of individual quasars can be derived from measurements of the proximity effect \citep[e.g.,][]{Khrykin_etal_2019,Eilers_etal_2017,Eilers_etal_2021}; these measurements often imply much shorter (e.g., $\lesssim 10^4$\,yrs) quasar lifetimes, suggesting quasar accretion is episodic in nature. 

In this work, we use a simple statistical argument to constrain the episodic quasar lifetime. Quasar accretion events, even if episodic, occur on much longer timescales than human lifetime. For example, for steady-state quasar accretion disks described by the $\alpha$-disk model \citep{Shakura_Sunyaev_1973}, the viscous infall timescale over which the accretion rate can change significantly across the UV-optical emitting part of the disk is on the order of centuries for a $10^8\,M_\odot$ SMBH. However, if we observe a large number of quasars over an extended observing window, which samples random phases in one episodic accretion event of these quasars, probabilistically we will find some of these quasars turning off during this observing period. The fraction $f_{\rm TO}$ of turned-off quasars over this period $\Delta t$ can then be used to estimate the average episodic lifetime $t_{\rm ep}=\Delta t/f_{\rm TO}$. This idea was first introduced in \citet{Martini_Schneider_2003}, and here we develop it further with recent findings from long-term quasar monitoring observations as well as simulations of quasar accretion disks. For simplicity, here we assume a constant episodic lifetime for all quasars; otherwise $t_{\rm ep}$ is averaged over the population of quasars. 

This paper is organized as follows. In \S\ref{sec:method} we detail our methodology and results, introducing an additional statistic related to recently turned-off quasars. We discuss caveats and implications of our results in \S\ref{sec:disc} and conclude in \S\ref{sec:con}. All timescales refer to the quasar rest-frame unless otherwise specified. The term ``quasar'' refers to unobscured broad-line quasars of $>10^8\,M_\odot$ SMBHs accreting at an Eddington ratio of $L_{\rm bol}/L_{\rm Edd}\sim 0.1$, where $L_{\rm bol}$ and $L_{\rm Edd}$ are the bolometric and Eddington luminosities of the quasar. This definition corresponds to a nominal luminosity threshold of $L_{\rm bol}>10^{45}\,{\rm erg\,s^{-1}}$ (or $M_i<-22$) for quasars. Throughout this work we use the phrases ``turned-off'' and ``turned-on'' to refer to events where the accretion rate significantly deviates from the prior steady state, accompanied by large flux changes. 

\section{Methodology and Results}\label{sec:method}

Quasars are not constant ``light bulbs'', and they display stochastic variability over the full ranges of timescales and wavelengths. In the UV-optical emission from the accretion disk, quasar variability typically reaches $\sim 10-20\%$ on months-to-years timescales \citep[e.g.,][]{Suberlak_etal_2021}, with a small fraction ($\sim 10\%$) of them reaching as much as one magnitude on multi-year timescales \citep[e.g.,][]{MacLeod_etal_2016,Rumbaugh_etal_2018}. There are now mounting evidence that most of these observed extreme variability quasars are not due to transient events such as dust obscuration, microlensing, or tidal disruption events \citep[e.g.,][]{Lawrence_2018}. Such evidence comes from spectral analysis of the bright and dim phases \citep[e.g.,][]{Runnoe_etal_2016}, polarimetric measurements \citep[e.g.,][]{Hutsemekers_etal_2017}, and observations of delayed echoes in the mid-infrared dust emission \citep[e.g.,][]{Sheng_etal_2017}. Furthermore, \citet{Rumbaugh_etal_2018} showed that there are systematic differences between extreme variability quasars and normal quasars, indicating that intrinsic accretion processes is the origin for the observed extreme variability. By extension, the so-called ``changing-look'' or ``changing-state'' quasars \citep[e.g.,][]{LaMassa_etal_2015,Yang_etal_2018,Graham_etal_2020}, the most extreme events where a quasar apparently changes spectral types between Type 1 and Type 2, are mostly recently turned-off or turned-on quasars. 

The terms ``extreme variability'', ``changing-look'' and ``changing-state'' have different definitions and context in different studies. \citet{Rumbaugh_etal_2018} defined extreme variability quasars as those with maximum variability of 1 magnitude in $g$ band, which is a simple and reproducible definition. ``Changing-look'' quasars are often defined as those with disappearing/appearing broad Balmer emission, but this definition depends on the signal-to-noise ratio of the faint-state spectrum. A more severe problem of this definition of changing-look quasars is that many of them have persistent broad \MgII\ emission even in the faint state \citep[e.g.,][]{MacLeod_etal_2019,Yang_etal_2020a}. \citet{Graham_etal_2020} suggested the name ``changing-state'' quasars to refer to those normally called ``changing-look'' quasars, but the ambiguity remains as what corresponds to a state transition. In this work, we use a simple, quantitative and reproducible definition of state transitions with specific applications to turn-off quasars (i.e., dimming by $>2$ mag in $g$), as detailed below.

The ubiquitous luminosity fluctuations of quasars do not necessarily mean that the accretion disk is not in a steady state in terms of the mass accretion rate. In recent global radiative MHD simulations of quasar accretion disks \citep[e.g.,][]{Jiang_etal_2019,Jiang_etal_2020}, light curve variability is produced even when the mass accretion rate at large radii is roughly constant. These simulated accretion disk light curves have variability amplitudes (between high and low flux points) as large as a factor of $\sim 2$ on multi-year timescales \citep{Jiang_etal_2019}. In extreme cases where a certain region of the inner (UV-emitting) accretion disk is convectively unstable due to enhanced iron opacity, the simulated light curve fluctuations can reach as much as a factor of $\sim 3-6$ on multi-year timescales \citep{Jiang_etal_2020}. Given high computational demand, these simulations have only been performed for limited cases. Nevertheless, these recent simulation results offer theoretical insights on episodes of quasar accretion.   

Since quasars are variable even in steady-state accretion, it is a non-trivial task to define a quasar episode and its ``turn-off''. Our following discussions are based on our best attempt to define quasar episodes in a reasonable and quantitative way, but the reader should keep in mind the ambiguity in the definitions. In \citet{Martini_Schneider_2003}, the definition of the turn-off of a quasar episode is that the quasar luminosity drops below $M_i=-23$ after accounting for typical quasar variability. This is a reasonable choice. However, a significant drop in the accretion rate (i.e., the ending of one episode) of a luminous quasar may still keep the accreting SMBH in the quasar regime, but the accretion state has changed from the prior steady state. In reality, the quasar may accrete at various rates at different times and never go completely quiescent, and one can always define an episode by requiring the accretion luminosity to be within a specific factor of the mean luminosity. 

In this work, we define a quasar episode as the phase during which the accretion is approximately in a steady state in terms of the mass loading rate at large disk radii, and the ``turn-off'' of this episode refers to when the quasar continuum dims by 2 magnitude in $g$ band, i.e., a transition of the accretion state or accretion event. $g$ band is preferred over longer wavelength bands given less contamination from host light. This definition of the turn-off of one episode is of course somewhat arbitrary, but does not undermine our reasoning. It provides a quantitative definition based on observations, and is motivated theoretically by recent quasar accretion disk simulations \citep{Jiang_etal_2019,Jiang_etal_2020}, as elaborated below. 

\begin{figure}
  \centering
    \includegraphics[width=0.48\textwidth]{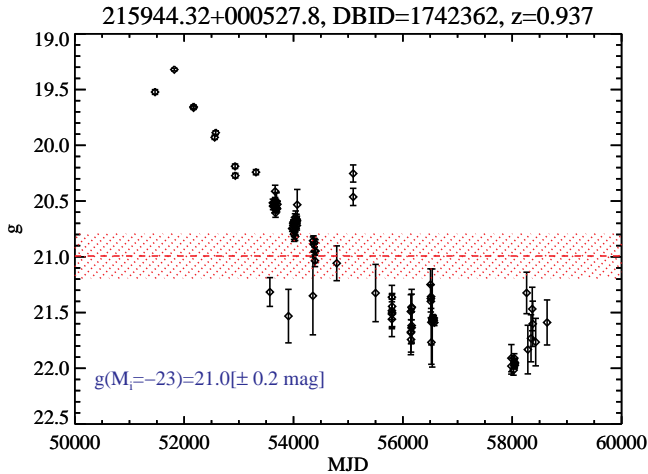}
    \caption{$g$-band light curve of an example turned-off quasar at $z=0.937$ in our sample. The light curve data are taken from \citet{Yang_etal_2020b} that extend beyond the baseline in \citet{Rumbaugh_etal_2018}. The red dashed line and band indicate the $g$-band magnitude and $\pm 0.2$ mag range corresponding to the definition of quasars of $M_{i}=-23$ in \citet{Martini_Schneider_2003}, which is one magnitude brighter than our definition of quasars. The full figure set of the 24 turned-off quasars with available light curves from \citet{Yang_etal_2020b} is provide as an online figure set. }
    \label{fig:examp}
\end{figure}

%Observationally, we define ``turn-off'' of an episodic accretion event when the quasar dims in continuum luminosity by 2 magnitude in $g$ band. This definition is of course somewhat arbitrary, but does not undermine our reasoning. If we defined a less stringent threshold for the turn-off, e.g., dimming by one magnitude, we would have a larger fraction of such quasars, and therefore a shorter average episodic lifetime. 

Using a dimming magnitude of 2 (a factor of $\sim 6$ in flux) is a reasonable choice, since it is too large to be consistent with steady-state accretion in latest simulations; it would also produce less ambiguous ``state transitions'' when observing the quasar spectroscopically at the bright and faint states in the presence of the host light. A smaller dimming magnitude (e.g., 1 magnitude) does not lead to unambiguous state transitions from an observational point of view. In addition, luminosity fluctuations of 1 magnitude can be produced on multi-year timescales in simulations of quasar accretion disks roughly in a steady state \citep{Jiang_etal_2019,Jiang_etal_2020}. On the other hand, although using a larger dimming magnitude would ensure we capture the turn-off of a quasar episode, it would result in poor statistics to constrain the fraction of turned-off quasars (e.g., only a handful of quasars would be classified as turn-off quasars in our sample below). More importantly, using a larger dimming magnitude would miss some genuine turn-off quasars, since the presence of host emission would necessarily limit the dynamic range of possible magnitude dimming for the whole system. For example, using spectral decomposition, \citet{Vandenberk_etal_2006} found $\sim 60\%$ of SDSS quasars ($M_i<-22$) at $z<0.75$ have a host-to-total fraction $f_H>0.1$ at rest-frame $\sim 4200$\,\AA, with a median host fraction of 0.3. A host fraction of 0.3 (0.1) would limit the maximum dimming to be 1.3 (2.5) magnitude. 

There is another empirical justification for our definition of turn-off quasars. A dimming factor of 2 magnitude roughly reduces the nominal Eddington ratio of $\sim 0.1$ to $\sim 0.01$, where there is a change in the behavior of the X-ray-to-optical ratio $\alpha_{\rm ox}$, indicative of a qualitative change in the accretion flow structure that signals a state transition \citep[e.g., figure 8 in][]{Jin_etal_2021}.

A factor of $\sim 6$ change ($\Delta g=2$) in the quasar continuum flux is sufficient to produce the apparent appearance/disappearance of the broad-line emission powered by the quasar continuum, leading to the observational identification of ``changing-look'' quasars, although the latter term is not rigorously defined. If we monitor a large sample of quasars over an extended period, the fraction of turned-off quasars within this period among the general quasar population constrains the episodic lifetime, regardless of how many episodes a quasar experiences during its entire cumulative lifetime.  

To apply this statistic, one can start from a well-defined quasar sample, and count the number of turned-off quasars over a fixed time window. Here we use  spectroscopically confirmed quasars identified from the SDSS in the Stripe 82 region and re-observed in the Dark Energy Survey (DES) imaging survey \citep[][]{Shen_etal_2011,Rumbaugh_etal_2018}. The much deeper DES photometry ensures all SDSS quasars are detected in the later DES epochs even when the quasar turns off, leaving only the host galaxy emission. The combined SDSS+DES optical light curves span $\sim 16$\,yrs in the observed frame \citep[][]{Rumbaugh_etal_2018,Luo_etal_2020}. 

There are 7,481 SDSS Stripe 82 quasars covered in DES with decade-long light curves \citep{Rumbaugh_etal_2018}, which is our parent sample. Although a larger photometric quasar sample is available \citep{Peters_etal_2015} in the Stripe 82 region, most of them do not have accurate spectroscopic redshifts, and are too faint such that host contamination would cause incompleteness in the identification of turn-off quasars. The statistics of extreme variability quasars from the systematic search in \citet[][]{Rumbaugh_etal_2018} are directly used for our selection of turned-off quasars.

Among these 7,481 quasars, there are 25 quasars that turned off (dimmed by more than 2 magnitude in $g$) within this observing baseline. We further exclude one quasar (with negligible impact on our statistics) that has no available optical light curve from \citet{Yang_etal_2020b}. The fraction of turned-off quasars is thus $\sim 0.32\pm0.06\%$ (24/7481), where uncertainties are estimated using Poisson statistics (bootstrapping error bars are similar). The quasar rest-frame duration of these observations is $\sim 8$\,yrs averaged over the redshift distribution of the sample (median $z\sim 1.2$). Therefore we estimate an average episodic quasar lifetime of $t_{\rm ep}\sim 2500\pm 470$ yrs. The true uncertainties are obviously larger, given the ambiguity of defining state transitions, but unlikely to change the episodic lifetime by an order of magnitude. 

We examined the optical light curves of these 24 turned-off quasars compiled by \citet[][]{Yang_etal_2020b}. About half of them show gradual fading over $\sim 16$ years which are generally difficult to reproduce with stochastic variability \citep{Luo_etal_2020}, while the remainder may recover their luminosity to some extent near the end of the light curve. Some of these quasars seem to re-brighten significantly within a few years after reaching the faintest state; we retain such objects in our turn-off quasar sample, and discuss their implications in \S\ref{sec:disc}. Two of them, however, show rapid (weeks to months) and large-amplitude optical variations likely dominated by blazar activity. Nevertheless, this fraction (2/24) is too small to affect our statistical estimation. Likewise, the fraction of radio-loud quasars in the sample is also too small to impact our results significantly. One example light curve is shown in Fig.~\ref{fig:examp}, and the full figure set is available as an online figure set. Table~\ref{tab:sample} lists the basic information of these 24 turned-off quasars.

%\url{http://quasar.astro.illinois.edu/paper_data/qso_lightcurve/lc_plots.pdf}

%\red{Maybe exclude those that re-brighten to previous levels. Maybe not. } 

\begin{table*}
\caption{Summary of the 24 turn-off quasars from the Stripe 82+DES sample in \citet{Rumbaugh_etal_2018}. Col (1) lists the object ID (DBID) used in the light curve catalog of \citet{Yang_etal_2020b}. Col (4) indicates if this object was detected in the FIRST radio survey (1=detection; 0=non-detection; $-1$=not covered by FIRST) from the catalog in \citet{Shen_etal_2011}. }\label{tab:sample}
%One object (DBID=2092378) has no corresponding light curve data in \citet{Yang_etal_2020b} even though it has a valid DBID.
\centering
%\scalebox{1.0}{
\begin{tabular}{cccc}
\hline\hline
DBID & J2000 designation &  Redshift & FIRST detection  \\
(1) & (2)  & (3) & (4)  \\
\hline
  68411 & 001130.40$+$005751.7 & 1.4934 & 1 \\ 
  88455 & 001420.44$-$003620.3 & 0.9589 & 0 \\ 
1332516 & 014048.62$-$003500.9 & 1.3839 & 0 \\ 
1222002 & 015335.63$+$010353.5 & 1.6176 & 0 \\ 
1667120 & 020537.13$+$002644.3 & 1.5136 & 0 \\ 
1647982 & 021752.75$+$001521.6 & 1.1805 & 0 \\ 
2024892 & 022252.26$-$000942.4 & 0.8222 & 1 \\ 
%     $-1$ & 022449.02$-$003935.4 & 0.5439 & 0 \\ 
2031395 & 022533.10$+$000308.3 & 2.2563 & 0 \\ 
2121523 & 022652.24$-$003916.5 & 0.6252 & 0 \\ 
%2092378 & 023230.22$+$004639.6 & 2.0653 & 0 \\ 
2046931 & 023814.15$+$001941.4 & 1.6023 & 0 \\ 
2551347 & 024741.84$+$011335.2 & 0.5396 & 0 \\ 
2474612 & 024837.54$+$011119.2 & 0.8088 & 0 \\ 
2597579 & 025505.68$+$002522.9 & 0.3535 & 0 \\ 
2582415 & 025515.09$+$003740.5 & 1.0228 & 1 \\ 
2564935 & 025712.75$-$004253.8 & 1.5716 & 0 \\ 
2655567 & 211817.39$+$001316.7 & 0.4629 & $-1$ \\ 
%     $-1$ & 215610.32$+$001753.2 & 0.5246 & 0 \\ 
1742362 & 215944.32$+$000527.8 & 0.9367 & 0 \\ 
1571037 & 221302.57$+$003015.9 & 0.7614 & 0 \\ 
%     $-1$ & 223734.15$+$011034.8 & 0.5233 & 1 \\ 
 779102 & 224433.56$+$000735.4 & 1.5434 & 0 \\ 
 861520 & 224924.01$+$004750.4 & 1.3500 & 1 \\ 
 281099 & 232838.04$-$005313.6 & 1.5531 & 0 \\ 
3935573 & 234142.32$+$003312.4 & 0.9861 & 1 \\ 
3950342 & 234855.04$+$002539.1 & 1.2752 & 0 \\ 
3924967 & 235936.81$-$003112.7 & 1.0956 & 1 \\ 
\hline
\hline\\
\end{tabular}
%}
\end{table*}

There is a second statistic related to recently turned-off quasars, based on observations of the galaxy rather than the quasar population. \citet{Roig_etal_2014} discovered a population of massive red galaxies that display broad \MgII\ emission on top of mainly a passive galaxy spectrum (i.e., no quasar continuum) -- these systems are referred to as ``broad \MgII\ emitters''. Their narrow-line emission on average places these objects in the AGN branch on the BPT diagram \citep{BPT}. \citet{Guo_etal_2020} used photoionization calculations to argue that these broad \MgII\ emitters are recently turned-off quasars, where the broad \MgII\ emission is slowly responding to the faded quasar continuum, as determined by atomic physics and radiative transfer. Indeed, the broad \MgII\ emission will disappear over multiple years \citep[e.g.,][Hall~et~al., in prep]{Guo_etal_2019b}, corroborating the idea that the persisting broad \MgII\ is echoing the recently diminished quasar light\footnote{An analogy is light echoes from extended narrow-line regions (NLRs) in recently turned-off quasars, e.g., Hanny's Voorwerp \citep{Keel_etal_2012}. Such echoes can be long-lasting (depending on how far away they are from the faded central quasar), e.g., $\sim 10^4-10^5\,$yr after the death of the quasar \citep{Keel_etal_2012}, meaning they are much more common than broad \MgII\ echoes. The NLR echoes as a chronometer thus have a much more coarse temporal resolution than broad \MgII\ echoes. }. 

This temporary phase with broad \MgII\ emission but no quasar continuum provides a chronometer of the last stage of a quasar episode. This phase must be longer than $\sim$\ a few yrs, given the persistence of such broad \MgII\ emission \citep[e.g.,][Hall~et~al., in prep]{Guo_etal_2019b}. It must also be less than several decades, since the broad \MgII\ emission originates from the broad-line region, and photoionization calculations suggest it cannot sustain its strength when the central engine is gone for decades \citep{Guo_etal_2020}. Therefore a reasonable estimate of the average duration of this phase is $\Delta t_{\rm MgII}\sim 10\,$yrs (quasar rest-frame). The relative abundance of broad \MgII\ emitters and quasars combined with $\Delta t_{\rm MgII}$ then constrains the episodic quasar lifetime. 

Both \citet{Roig_etal_2014} and \citet{Guo_etal_2019b} reported the fraction of broad \MgII\ emitters among SDSS galaxies at $0.4\lesssim z\lesssim 0.8$. Given the flux limit of SDSS, these galaxies are mostly massive galaxies with stellar masses of $\sim 10^{11}\,M_\odot$ \citep{Chen_etal_2012}, and they are roughly the hosts for quasars with black hole mass of $\sim 10^8-10^9\,M_\odot$ within this redshift range. The selection generally requires well-detected broad \MgII\ emission and the absence of quasar-like blue continuum and broad \hbeta\ emission. But the details in the parent samples and the selection criteria are different in these two studies, resulting in slightly different sample statistics. \citet{Roig_etal_2014} reported a broad \MgII\ emitter fraction of $<0.1\%$ while \citet{Guo_etal_2019b} reported a fraction of $\sim 0.02\%$. In this work we use these two estimates to bracket the range of the fraction of broad \MgII\ emitters among massive galaxies at $z\sim 0.6$. 

The fraction of quasars among massive galaxies (or the duty cycle) at $z\sim 0.6$ is $\sim 1\%$ inferred from quasar clustering measurements \citep[e.g.,][]{Laurent_etal_2017}. Based on the relative frequency of broad \MgII\ emitters and quasars among massive galaxies, we estimate an average episodic quasar lifetime at $z\sim 0.6$ of $\sim 100-500\,$yr. 

%This $\sim 1\%$ quasar fraction is roughly in line with the fraction of Type 2 AGN\footnote{SMBHs accreting at substantially lower accretion rates than quasars (i.e., LINERs) are not included in the estimation of this fraction. In addition, the parent population is all SDSS galaxies, most of which do not have measurable narrow emission lines.} among SDSS galaxies at $0.4\lesssim z\lesssim 0.8$ \citep{Thomas_etal_2013}. 

The estimates of the episodic quasar lifetime from these two turn-off statistics differ by an order of magnitude, which we attribute to the difficulties of precisely matching the broad \MgII\ emitter and quasar populations among massive galaxies, as well as consistently defining the turn-off signatures for both statistics. Indeed, the original AGN luminosity of these broad \MgII\ emitters, traced by the \OIII\ luminosity\footnote{The typical effective radius of the \OIII\ surface brightness profile in Seyferts is tens to hundreds of parsec \citep[e.g.,][]{Schmitt_etal_2003}. For quasars, the \OIII\ emitting regions are larger, e.g., hundreds to thousands of parsec. It is possible that the innermost \OIII\ emitting region within $\sim 10\,$parsec has started to fade in response to the reduction in quasar flux, but it is unlikely that the bulk of the \OIII\ emission has faded given the typical narrow-line region size in quasars.}, is systematically lower than that of SDSS quasars by $\sim 0.8$\,dex (H.~Guo, private communications). Roughly a third of the \MgII\ emitters would have original AGN luminosities above the threshold of quasars ($L_{\rm bol}>10^{45}\,{\rm erg\,s^{-1}}$). Thus if we require the \MgII\ emitters to have quasar luminosities before turn-off, the estimated episodic lifetime will be revised up by a factor of $\sim 3$. 

Furthermore, a less extreme drop in the quasar continuum, e.g., by a factor of $\sim 2-3$, may produce ``orphan'' broad \MgII\ emission in a galaxy spectrum as well, because the reduced (but not turned-off) quasar continuum (and optical broad emission lines) may be difficult to detect in the presence of host galaxy light. In this case, the reported \MgII\ emitter fraction would significantly over-predict the incidents of turned-off quasars (defined as those faded by $>2$ mags), and underestimate the episodic lifetime. More dedicated photoionization calculations and better statistics of broad \MgII\ emitters are needed to fully address this discrepancy. Finally, the two turn-off quasar samples are not matched in redshift, which the episodic lifetime may depend on. The current sample statistics are insufficient to match these two samples in the redshift-luminosity space. 

%\red{Check to see if the broad \MgII\ emitters have systematically lower [OIII] luminosities than the SDSS+DES quasar sample. }

%if a slightly smaller drop in quasar flux than the 2 magnitude threshold we adopted can also produce a broad \MgII\ emitter, the episodic lifetime estimated from the first statistic would be reduced and brought closer to that estimated from the broad \MgII\ emitter statistic. \red{Alternatively, this could mean that some broad \MgII\ emitters are not truly turned-off quasars, but because of a factor of 2-3 reduction in quasar continuum. }

\section{Discussion}\label{sec:disc}

%\red{The assumption that a 2 magnitude dimming marks the transition is a little on shaky ground, and it can go either up or down. But using the two statistics helps to converge on a few thousand yrs of the episodic lifetime. }

Given that all the statistics we utilize are only accurate to a factor of few, we cannot place a precise constraint on the average episodic quasar lifetime. It is also well expected that the duration of quasar episodes would differ in different systems, at different times for the same system, or at different redshifts. However, from a statistical point of view, it is unlikely that the ensemble-averaged episodic lifetime is much longer than $10^4\,$yrs, because then we would observe a much smaller (by at least an order of magnitude) fraction of turned-off quasars with accumulated data in the last few decades -- essentially all quasars would be accreting and emitting more constantly over this observing window. 

The cumulative quasar lifetime is $\sim 10^6-10^8\,$yrs constrained from quasar clustering measurements and SMBH demographics (see \S\ref{sec:introduction}). It is unlikely that a SMBH will acquire most of its mass by accreting for $\lesssim 10^6$\,yrs at a moderate Eddington ratio of 0.1 and a radiative efficiency of 10\%, since the $e$-folding time is $5\times 10^8\,$yrs. Our results thus imply that during the assembly history of a SMBH, it must experience $\sim 10^3-10^5$ such relatively short quasar episodes in order to grow to its observed mass by accretion. Additional accretion may occur during the obscured phase, although the accreted SMBH mass density is already substantial during the unobscured phase as constrained from quasar clustering and SMBH demographics. 

However, these quasar episodes can either be distributed sparsely or highly clustered over the entire life span of a quasar. In gas-rich nuclear environment, the events that the SMBH replenishes its accretion disk may occur quite frequently, leading to short gaps in individual quasar episodes. For example, roughly a third of the turned-off quasars in our SDSS+DES sample seem to re-brighten (to some extent) in only a few years after dimming to the lowest flux state. These re-brightening events could be due to normal accretion disk variability at the fainter state, or signal the onset of another quasar episode. In the latter case, this rising phase has not yet reached steady state, thus its short rising time (e.g., a few years) is not in direct conflict with the much longer viscous time. This may explain why some proximity-effect measurements suggested short ($<10^4\,$yrs) quasar lifetimes while others suggested much longer lifetimes \citep[e.g.,][]{Khrykin_etal_2019,Eilers_etal_2021}: when the accretion disk is rapidly replenished, the proximity zone in the IGM around the quasar will feel no difference as if the quasar were continuously on. The same argument applies to the $\sim$~few kpc sizes of the quasar narrow-line region that require clustered quasar episodes over $\gtrsim 10^4\,$yrs \citep{Martini_2004,Schawinski_etal_2015}. 

If efficient mass accretion onto the SMBH is triggered in a major fueling event, e.g., from a gas-rich major merger, then it is possible that quasar episodes are grouped within periods of bursts that can last as long as $\sim 10^6\,$yrs, until stellar and/or AGN feedback interrupts the gas supply. For example, models of quasar-level accretion fueled by recycled gas in elliptical galaxies \citep{Ciotti_Ostriker_2007} predict bursts of $\sim 0.1-1\,$Myr in duration, separated by much longer gaps of quiescence due to AGN feedback. Similarly, cosmological hydrodynamic simulations of quasar-hosting galaxies show $\lesssim 1\,$Myr sporadic quasar bursts during the early growth of the SMBH, regulated by stellar feedback continuously evacuating gas from the galactic nucleus \citep[][]{Angles_etal_2017}. Of course, none of these galaxy simulations can resolve the quasar accretion disk, which is more directly related to the episodic quasar lifetime.

Smaller accreting SMBHs with luminosities in the Seyfert regime (i.e., $L_{\rm bol}\lesssim 10^{45}\,{\rm erg\,s^{-1}}$) could have shorter episodic lifetimes than quasars, since the relevant timescales (e.g., the viscous time) pertaining to accretion disks are generally shorter for smaller SMBHs. Indeed, monitoring of low-redshift Seyferts in the past few decades has discovered several ``changing-look'' AGNs that changed their spectral types back and forth frequently. One notable example is NGC 4151 (also Mrk 1018 and Mrk 590), where the broad emission lines had disappeared and later reappeared over the course of several decades \citep[e.g.,][]{Osterbrock_1977,Penston_Perez_1984,Shapovalova_etal_2010}. The frequency of such dramatic flux changes is seemingly higher in Seyferts than in quasars. The implied episodic accretion time is then only tens of years for Seyferts. 

%\red{Mention Seyferts (presumably the episodic lifetime is shorter); mention the viscous time; mention obscured quasars and how that fits into the picture. }

While we have focused on recently turned-off quasars, in principle the statistics of recently turned-on quasars can constrain the episodic lifetime as well. This can either be done with historic data for a quasar sample constructed from a later survey \citep[e.g.,][]{Martini_Schneider_2003}, or observing a galaxy sample compiled in an earlier survey at later times. The latter approach requires a well characterized parent galaxy sample (basically, we require an estimate of the space density of recently turned-on quasars). Moreover, some of these ``turned-on'' events are transient tidal disruption events in a fully grown SMBH, which are different from the regular feeding cycles during the major growth period of the SMBH. Nevertheless, it is worth exploring such galaxy-based statistics with multi-epoch observations and careful matching between different samples, e.g., redshift, quasar luminosity, etc. \citep[e.g.,][]{Yang_etal_2018}.

Finally, we comment on the comparison with the results in \citet{Martini_Schneider_2003}. Using SDSS Early Data Release (EDR) quasars and historic POSS imaging, \citet{Martini_Schneider_2003} derived a lower limit of $\sim 20,000\,$yr for the episodic quasar lifetime, seemingly much longer than our estimates. Despite the $\sim 50$ yr baseline between POSS and SDSS, they did not find any SDSS EDR quasars that were not ``on'' at the POSS epoch, after accounting for the detection limit in POSS. It is possible that the POSS+SDSS constraints only apply to the most luminous quasars, which may have a longer average episodic lifetime than moderate-luminosity quasars as discussed earlier. If we use the same definition as in \citet{Martini_Schneider_2003} for the ``on'' phase, $M_i<-23$, and require that the quasar stays ``off'' after the dimming (otherwise the temporary state transition may be missed by the two-epoch POSS+SDSS observations), there are still $\sim 10$ ``turned-off'' quasars in our SDSS+DES sample, resulting in an average episodic lifetime of $\sim 7000\,$yr. 

\section{Conclusions}\label{sec:con}

Variability is one of the defining characteristics of quasars. However, it has only become well recognized recently that there are a small fraction of quasars that show extreme variability over years to decades timescales, indicative of transitions in their accretion state. The samples of quasars and galaxies with extended temporal coverages are now large enough to derive meaningful constraints on the average episodic lifetime of quasar activity. The longer quasars accrete at more or less steady rates (with luminosity fluctuations within a factor of few), the less frequently we will observe quasars in ``state transitions''. 

In this work we used the statistics of recently turned-off quasars to constrain the episodic quasar lifetime. Using two statistics, the fraction of turned-off quasars within a given observing baseline and the relative frequency of broad \MgII\ emitters in galaxies with respect to quasars, we constrain the average episodic quasar lifetime to be hundreds to thousands of years. These order-of-magnitude constraints are mainly limited by the difficulty of precisely defining the state transition (for the first statistic), and uncertainties in the relative frequency of broad \MgII\ emitters and quasars among the galaxy population (for the second statistic). Continued time-domain observations of quasars and galaxies in the coming decades will be able to improve these constraints. In particular, we expect the sample statistics of turn-off quasars will improve both due to longer observing baselines and larger parent samples. These improved sample statistics will allow us to explore the redshift and quasar luminosity dependences of the episodic lifetime.

%\red{Explain how to improve, and mention matching redshift, luminosity, etc. with better statistics. }

Nevertheless, our simple statistical arguments strongly disfavor average episodic quasar lifetimes of $>10^4\,$yrs. Such long, steady (luminosity within a factor of few) accretion will produce much less frequent incidents of recently turned-off quasars than observed. On the other hand, because the majority of quasars have been observed to be more or less stable over the past $\sim 60\,$yrs since the discovery of quasars \citep{Schmidt_1963}, quasar episodes must be $>100\,$yrs on average. It is remarkable that our estimates of the episodic quasar lifetime are broadly consistent with the accretion disk viscous infall time (at $\sim 100R_g$) for $10^8-10^9\,M_\odot$ SMBHs, which suggests that quasar episodes are governed by fundamental accretion disk physics, rather than by the gas supply at much larger spatial scales.

Finally, ``changing-look'' or ``changing-state'' quasars are commonly deemed difficult to explain given the long (hundreds to thousands of years) viscous time of steady-state accretion disks and the much shorter timescales (years to decades) over which the transition is observed \citep[e.g.,][]{Lawrence_2018}. We point out that this might be a misconception: if quasars significantly change their accretion states over centuries on average, then it is guaranteed that we will catch a small fraction of them transitioning within much shorter (e.g., years to decades) observing windows\footnote{Alternatively, one may argue that changing-look quasars are a rare sub-class of quasars that have nothing to do with episodic quasar accretion, which would then require modified accretion disk theories. But Occam's razor would favor our simple statistical interpretation of changing-look quasars, which predicts an increasing frequency of such objects as the observing baseline increases.}. It is possible, however, that when approaching this ``rising/fading'' phase of the quasar episode, the accretion flow becomes less stable, resulting in larger luminosity fluctuations on days to years timescales than those in the steady accretion phase, as observed \citep{Rumbaugh_etal_2018,MacLeod_etal_2019}. In addition, the transitions themselves could complete in a relatively quick fashion ($\lesssim$ a few years), during which the accretion is unlikely to be in a steady state.  

%\red{If the episode lifetime is very long, and all CLQs are a special subclass of quasars, then we should not see the CLQ fraction increases when the observing baseline increases, assuming the baseline is long enough to capture all these special quasars. }

%One may still argue that ``changing-look'' quasars are a special class of accreting SMBHs that defies current accretion theories, but Occam's razor would favor the simpler interpretation. 

\acknowledgments

We thank the referee for useful comments that improved the manuscript, and Paul Martini, Hengxiao Guo, Qian Yang, Mike Eracleous, Yan-Fei Jiang, Xin Liu, Scott Anderson, and Paul Green for useful comments and discussions. YS acknowledges support from NSF grant AST-2009947. 

% Yan-Fei Jiang, Xin Liu, Hengxiao Guo, Qian Yang, Mike Eracleous

\bibliography{qso_lifetime_rv1.bbl}

\end{document}